\documentclass[conference]{IEEEtran}
\ifCLASSINFOpdf
\else
\fi
\usepackage{soul}
\usepackage{authblk}
\usepackage{bm}
\usepackage{xcolor}
\usepackage{stackengine}
\def\delequal{\mathrel{\ensurestackMath{\stackon[1pt]{=}{\scriptstyle\Delta}}}}
\usepackage{cite}
\usepackage{amssymb}
\usepackage{mathtools}
\usepackage{mathrsfs}
\usepackage{eucal}
\usepackage{amsmath}
\usepackage{mathptmx}

\usepackage{graphicx}
\usepackage{epstopdf}
\usepackage{pgfplots}
\usepackage{tikz}
\usetikzlibrary{patterns}
\usepackage{array}
\usepackage{subfigure}

\graphicspath{ {./images/} }
\usepackage[utf8]{inputenc}
\usepackage{comment}
\usepackage[utf8]{inputenc}
\usepackage{multirow}
\usepackage{url}
\usepackage{multicol}
\usepackage{xcolor}
\usepackage[linesnumbered,ruled,vlined]{algorithm2e}
\usepackage{algorithmic}

\SetCommentSty{mycommfont}
\SetKwInput{KwInput}{Input}
\SetKwInput{KwInitialize}{Initialize}
\SetKwInput{KwOutput}{Output} 

\begin{document}
\setlength{\parskip}{0pt}
\title{A Novel Algorithm to Report CSI in MIMO-Based Wireless Networks}
\author[1, 2]{Muhammad Karam Shehzad}
\author[1]{Luca Rose}
\author[2]{Mohamad Assaad}
\affil[1]{Nokia Bell Labs\\France.}
\affil[2]{Laboratoire des Signaux et Systemes, CentraleSupelec, Gif-sur-Yvette, France.}
\affil[ ]{Emails: muhammad.shehzad@nokia.com, luca.rose@nokia-bell-labs.com, mohamad.assaad@centralesupelec.fr}
\maketitle

\begin{abstract}
In wireless communication, accurate channel state information (CSI) is of pivotal importance. In practice, due to processing and feedback delays, estimated CSI can be outdated, which can severely deteriorate the performance of the communication system. Besides, to feedback estimated CSI, a strong compression of the CSI, evaluated at the user equipment (UE), is performed to reduce the over-the-air (OTA) overhead. Such compression strongly reduces the precision of the estimated CSI, which ultimately impacts the performance of multiple-input multiple-output (MIMO) precoding. Motivated by such issues, we present a novel scalable idea of reporting CSI in wireless networks, which is applicable to both time-division duplex (TDD) and frequency-division duplex (FDD) systems. In particular, the novel approach introduces the use of a channel predictor function, e.g., Kalman filter (KF), at both ends of the communication system to predict CSI. Simulation-based results demonstrate that the novel approach reduces not only the channel mean-squared-error (MSE) but also the OTA overhead to feedback the estimated CSI when there is immense variation in the mobile radio channel. Besides, in the immobile radio channel, feedback can be eliminated, which brings the benefit of further reducing the OTA overhead. Additionally, the proposed method provides a significant signal-to-noise ratio (SNR) gain in both the channel conditions, i.e., highly mobile and immobile. 
\end{abstract}
\IEEEpeerreviewmaketitle
\section{Introduction} \label{sec1}
In the modern era of wireless communication, the unabated growth of cellular users and their immense data rate demands require novel advancements in the existing cellular infrastructure. To meet the requirements of cellular users, many transmission techniques such as bit-loading, coding, precoding methods, adaptive modulation, channel-aware scheduling, etc., are bounded to have accurate channel state information (CSI) at the transmitter side, to achieve a significant gain. Also, accurate CSI notably improves the performance of many wireless techniques, for instance, multiple-input multiple-output (MIMO) \cite{MIMO}, ultra-reliable transmissions \cite{URLLC}, relaying \cite{relay} and physical layer security \cite{security}. Moreover, accurate CSI reaps benefits of simplifying the receiver through MIMO precoding at the transmitter, reliability, and higher link capacity, etc. To acquire these gains, CSI at the transmitter is indispensable. Therefore, accurate CSI is, interestingly, the alpha and omega of modern wireless communication infrastructure.
\par Keeping in lieu the above benefits of accurate CSI, it is, however, challenging to obtain  precise CSI in fifth-generation (5G) environment, for instance, millimeter-Wave (mmWave), critical machine type communication (cMTC)-vehicular communication, factory automation, etc. While, in theory, time-division duplex (TDD) systems could exploit reciprocity to obtain CSI, this aspect is not fully exploited in current 5G networks, where the user equipments (UEs) feedback a compressed estimation of the downlink channel to the base station (BS). 
On the other hand, in frequency-division duplex (FDD) systems, the requirement of feedback increases with the increase in the number of transmit antennas; thereby, the design of MIMO system with limited feedback becomes a critical issue \cite{lim_fb}. In addition, currently, in 5G, two different strategies are considered for CSI report: type-I CSI and type-II CSI feedback \cite{3GPP1}, \cite{3GPP2}. Nonetheless, both methods involve a strong compression to reduce the over-the-air (OTA) overhead, which further reduces the precision of the CSI, consequently impacting the performance of MIMO precoding.  
\par Recently, machine learning (ML) techniques for wireless communications have attracted attention of researchers \cite{rose,hoydis}. In particular channel prediction, which is capable of forecasting the future CSI, has attracted an eye of researchers both from academia and industry. To this end, statistical methods \cite{AR1}, \cite{AR2}, which model the fading channel as an auto-regressive (AR) process and later employ Kalman filter (KF) to predict the future channel realizations, have been proposed in the literature. For example, \cite{KF1,KF2,AKF} use the KF as a channel predictor and estimator, for single-input single-output (SISO) and MIMO radio channels. Nevertheless, the aim of all the aforementioned work is to reduce the channel mean-squared-error (MSE); therefore, the practical deployment aspects of the KF have been ignored. Above all, feeding back the estimated CSI and its practical implication, e.g., compression, etc., have also been neglected, which are of predominant importance and are the cornerstone of cellular infrastructure. 
\begin{figure*} 
\begin{center}
  \includegraphics[width=18cm,height=8cm]{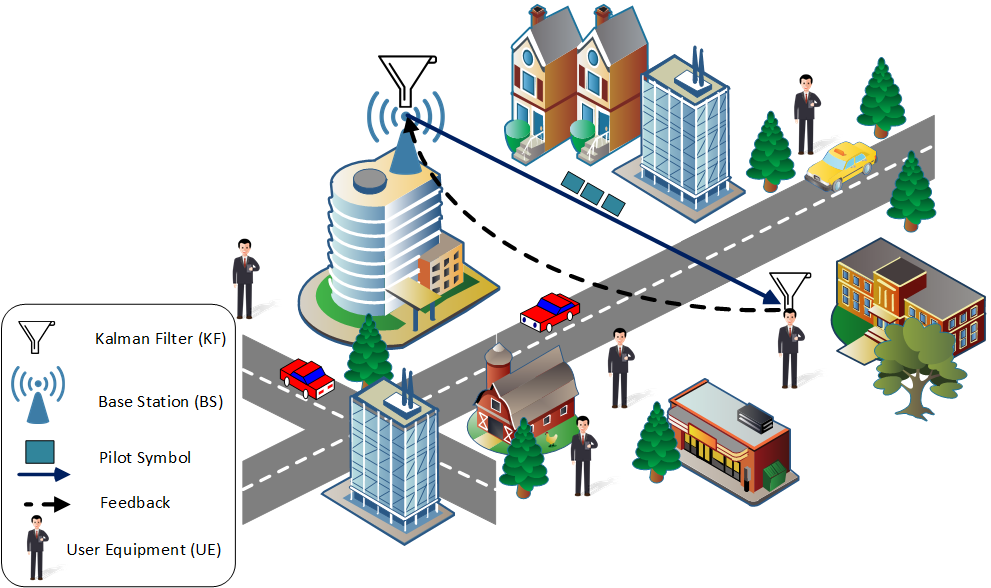}
  \caption{{KF-enabled communication scenario.}}\label{scenario}      
\end{center}
\end{figure*} 
\par In this paper, keeping in view the aforementioned issues, the proposed work concerns the utilization of a channel predictor function, i.e., KF, to help the CSI reporting function in 5G networks. Specifically, we propose a novel way to exploit KF to either reduce the quantization error (hence, improving MIMO precoding performance) while keeping the OTA occupation constant or to reduce the overhead further. In particular, the novel approach adopts KF at both ends of the communication system, i.e., BS and the UE. The predictor, running at both sides, exploits the same inputs in order to predict the next channel realization; consequently, providing the same prediction at both ends. The output of the KF will be used as a base, and an updated channel measure, CSI, will be fed back with respect to the prediction. As a toy example, if the prediction were perfect, and there was no difference between the predicted and measured CSI, then no feedback would be necessary.
\par The remainder of the paper is organized as follows. In Section\,\ref{systemmodel}, system model is introduced. Additionally, the dynamics of the time-varying radio channel are discussed in the same section. The novel method for reporting CSI, along with the advantages, is explained in Section\,\ref{proposed}. A detailed analysis of the proposed method is presented in Section\,\ref{sec4}. Finally, conclusions are drawn in Section\,\ref{conclusion}.  
\par $Notations:$ The transpose, conjugate, and Hermitian transpose are denoted by $(\cdot)^{T}$, $(\cdot)^{*}$, and $(\cdot)^{H}$, respectively. In addition, matrices are represented by boldface upper-case, vectors with boldface lower-case, and scalars with normal lower-case. Further, $\mathbf{0}_{p\times 1}$ and $\mathbf{I}_{p\times p}$ denote the null vector of dimension $p\times 1$ and identity matrix of dimension $p\times p$, respectively. Moreover, $h$, $\hat{h}$, and $\widetilde{h}$ represent the actual, estimated and predicted channel, respectively. 
\section{System Model}\label{systemmodel}
\subsection{Communication Environment}
Consider the downlink of a single-user MIMO (SU-MIMO) cellular system, as depicted in Fig.\,\ref{scenario}, in which two network entities, that is, UE and a BS are set. Further, a channel predictor function, i.e., KF, is deployed at both the network entities. As shown in Fig.\,\ref{scenario}, the UE served by the BS, estimates the channel by receiving dedicated reference symbols (RS), transmitted by the BS. Later on, the compressed version of the estimated channel is fed back to the BS by the UE. It is important to note that, for the sake of simplicity, we assume the communication between a single BS-UE link, however, the same process would be followed by the remaining UEs, present in the network. To this end, let us denote the BS and the UE by $q$ and $\varsigma$, respectively. In addition, $q$ and $\varsigma$ are equipped with ${i} = \{1, 2, 3, ..., N_{t}\}$ and ${j} = \{1, 2, 3, ..., N_{r}\}$ antennas, respectively. In the following, we discuss the signal model and channel dynamics for the communication scenario given in Fig.\,\ref{scenario}. 
\subsection{Signal Model}
Assuming a multicarrier transmission, $k$ and $n$ represent the subcarrier index and the time instant of a RS, respectively, where ${k} \delequal \{1, 2, 3, ..., K\}$. Therefore, from $i^{th}$ transmit antenna to $j^{th}$ receive antenna, the received signal, at the UE, from the $k^{th}$ subcarrier, at the $n^{th}$ time instant is written as 
\begin{equation} \label{rx signal}
\gamma^{j}_k(n)= h_k^{i,j}(n)\cdot s^{i}_k(n)+v_k^{j}(n)\:,
\end{equation}
where $s^{i}_k(n)$ is the information symbol of the $i^{th}$ transmit antenna, carried by the $k^{th}$ subcarrier at the $n^{th}$ time instant, and $h^{i,j}_k(n)$ is the channel gain. Additionally, $v^{j}_k(n)$ is the complex additive white Gaussian noise (AWGN) having mean zero and variance $\sigma_{v}^{2}$.     
\subsection{Channel Dynamics}
The AR process \cite{autoregressive}, is a powerful way of modeling the dynamic systems and has widely been used in the literature (e.g., \cite{KF1, KF2, AKF, AR1, AR2}) to predict the fading channel using KF-based channel estimation. Moreover, AR has also been considered to simulate correlated Rayleigh fading channels \cite{ray1,ray2}. Therefore, in our study, to model the dynamics of the time-varying radio channel, we also consider AR process. The complex AR process \cite{autoregressive} of order $p$, AR$(p)$, for the channel gain, $h^{i,j}_k(n)$, is given as
\begin{equation} \label{channel_AR}
h^{i,j}_k(n)=\sqrt{1-\tau^2}\sum_{l=1}^{{p}} c_k(n)\cdot h^{i,j}_k(n-l)+\tau\cdot w_k^{i,j}(n)\:,
\end{equation}
where $c_k(n)$ and $w^{i,j}_k(n)$ are the parameter of the AR process and the complex AWGN with zero-mean and variance $\sigma_{p}^{2}$, respectively. In addition, $\tau$ is a parameter that measures the one step correlation between the two consecutive channel realizations. In other words, $\tau$ measures how much the channel is predictable. The corresponding power spectral density (PSD) of the AR$(p)$ is
\begin{equation} \label{PSD}
P_c(f)=\frac{\sigma_{p}^{2}}{\left\lvert1+\sum\limits_{l=1}^{p} c_l\cdot e^{-j2\pi fl}\right\rvert^2}\:.
\end{equation}
The auto-correlation of the channel gain, $h^{i,j}_k(n)$, from the Jakes' fading model \cite{jakes1994microwave}, is written as
\begin{equation}\label{auto}
\begin{aligned}
R_k(l)&= E\big [h^{i,j}_k(n)\cdot h^{^{*} i,j}_k(n-l)  \big] \\
& = J_{0}(2\pi f_{m}l)\:, 
\end{aligned}
\end{equation}
where $f_{m}=f^{max}_{d}T_{s}$ represents the maximum Doppler shift, which is normalized by the sampling rate $f_{s}=\frac{1}{T_{s}}$. In addition, $J_{0}(\cdot)$ is the zeroth-order Bessel function of the first kind. Besides, by using the auto-correlation function given in \eqref{auto}, the AR model parameters can be obtained by solving Yule-Walker equation \cite{yule}, and these parameters are assumed to be known. Further, let us define
\begin{equation} \label{state}
\mathbf{x}^{i,j}_k(n)=\big [h^{i,j}_k(n), h^{i,j}_k(n-1), ..., h^{i,j}_k(n-p+1)\big ]^{T}\:,
\end{equation}
by using the above definition, the state equation is formulated as
\begin{equation} \label{state_eq}
\mathbf{x}^{i,j}_k(n)=\bm{\Phi}_{k} \mathbf{x}^{i,j}_k(n-1)+\mathbf{w}^{i,j}_k(n)\:,
\end{equation}
where $\bm{\Phi}_{k}$ and $\mathbf{w}^{i,j}_k(n)$ depict the state transition matrix and the state noise vector, respectively, and are written as follows
\[
\bm{\Phi}_{k}=
  \begin{bmatrix}
    c_{k}(1)&c_{k}(2)&\cdots & c_{k}(p-1)&c_{k}(p)\\
    1&0&\cdots&0&0\\
    0&1&\cdots&0&0\\
    \vdots&\vdots&\ddots&\vdots&\vdots\\
    0&0&\cdots&1&0&
  \end{bmatrix}
  \]
\begin{equation} \label{state noise}
\mathbf{w}^{i,j}_k(n)=\big [w^{i,j}_k(n), 0, ..., 0\big ]^{T}\:.
\end{equation}
Finally, the measurement equation or the observation equation of the state-space model is given below
\begin{equation} \label{measurement equation}
\mathbf{y}^{i,j}_k(n)=\mathbf{M}_k(n)\mathbf{x}^{i,j}_k(n)+\mathbf{v}^{i,j}_k(n)\:,
\end{equation}
where $\mathbf{M}_k(n)$ is the measurement matrix, and $\mathbf{v}^{i,j}_k(n)$ is the noise vector for the measurement equation, which is written as
\begin{equation} \label{measurement noise}
\mathbf{v}^{i,j}_k(n)=\big [v^{i,j}_k(n), 0, ..., 0\big ]^{T}\:.
\end{equation}
\par In the following section, we discuss the proposed method to report CSI.
\section{{Proposed Method}}\label{proposed}
The proposed method to report CSI utilizes a channel predictor based on KF, at both ends of the communication. Further, for each UE in the cell, the BS stores a certain amount of previous CSI estimations. This data, available at both ends, will be used by the predictors, to predict the next channel realization. For the sake of simplicity, we will describe the proposed method for a single BS-UE link, remarking that this process should be repeated for all the UEs supporting the channel predictor. 
\par Below, we first summarize the KF for channel estimation and prediction. Later on, we will describe the proposed method, which utilizes the KF.
\subsection{{Kalman Filter for Channel Estimation and Prediction}}\label{KF_prediction}
\par KF \cite{kalman1960new} is a Bayesian solution for estimating the state of a dynamic system, i.e., channel, in which, state evolution and measurement process are both linear. Broadly speaking, KF consists of two major steps, i.e., prediction and update. The former uses the state model to predict the state vector, ${\hat{\mathbf{x}}}^{i,j}_{k}(n|n-1)$, and the correlation matrix, $\mathbf{P}^{i,j}_{k}(n|n-1)$. Whereas, the latter updates the prediction, i.e., estimates the state vector, $\hat{\mathbf{x}}^{i,j}_{k}(n|n)$, and the correlation matrix, $\mathbf{P}^{i,j}_{k}(n|n)$, of the estimation error, based on the measurements $\mathbf{y}^{i,j}_k(1), \mathbf{y}^{i,j}_k(2), ..., \mathbf{y}_k^{i,j}(n)$. The KF is initialized as follows
\begin{equation}\label{KF_initilization}
\begin{aligned}
{\hat{\mathbf{x}}}^{i,j}_{k}(0|0)=\mathbf{0}_{p\times1} \\
{\mathbf{P}}^{i,j}_{k}(0|0)=\mathbf{I}_{p\times p}\:.
\end{aligned}
\end{equation}
Then, the KF calculates the following equations \cite{kalman1960new} recursively for each $n$.
\begin{itemize}
    \item Prediction step:
    \begin{equation} \label{eq:1}
{\hat{\mathbf{x}}}^{i,j}_{k}(n|n-1)=\bm{\Phi}_{k} {\hat{\mathbf{x}}}^{i,j}_{k}(n-1|n-1)
\end{equation}
\begin{equation} \label{eq:2}
\mathbf{P}^{i,j}_{k}({n|n-1})=\bm{\Phi}_{k} \mathbf{P}^{i,j}_{k}({n-1|n-1}) \bm{\Phi}^{H}_{k}+\mathbf{{Q}}_{w,k}(n)
\end{equation}
    \item Update step:
\end{itemize}

\begin{equation}
    \mathbf{G}^{i,j}_{k}(n)=\frac{\mathbf{P}^{i,j}_{k}(n|n-1) \mathbf{M}^{H}_{k}(n)}{\mathbf{M}_{k}(n) \mathbf{P}^{i,j}_{k}({n|n-1}) \mathbf{M}^{H}_{k}(n)+\mathbf{Q}_{v,k}(n)}
\end{equation}

\begin{equation} \label{eq:5}
\mathbf{e}^{i,j}_{k}(n)=\mathbf{y}^{i,j}_{k}(n)-\mathbf{M}_{k}(n) \hat{\mathbf{x}}^{i,j}_{k}(n|n-1)
\end{equation}
\begin{equation} \label{est_channel}
\hat{\mathbf{x}}^{i,j}_{k}(n|n)=\hat{\mathbf{x}}^{i,j}_{k}(n|n-1)+\mathbf{G}^{i,j}_{k}(n) \mathbf{e}^{i,j}_{k}(n)
\end{equation}
\begin{equation} \label{eq:7}
\mathbf{P}^{i,j}_{k}({n|n})=[\mathbf{I}_{p\times p}- \mathbf{G}^{i,j}_{k}(n) \mathbf{M}_{k}(n)] \mathbf{P}^{i,j}_{k}({n|n-1})
\end{equation}
\begin{equation} \label{pred_channel}
\widetilde{h}^{i,j}_k(n+1)=\big [\bm{\Phi}_{k}\hat{{\mathbf{x}}}^{i,j}_{k}(n|n)\big ]_{1}
\end{equation}
where
\begin{equation*}
    \mathbf{{Q}}_{w,k}(n)=E[\mathbf{w}^{i,j}_k(n) \mathbf{w}^{^{H} i,j}_k(n)]
\end{equation*}
\begin{equation*}
    \mathbf{{Q}}_{v,k}(n)=E[\mathbf{v}^{i,j}_k(n) \mathbf{v}^{^{H} i,j}_k(n)]\:,
\end{equation*}
$\mathbf{G}^{i,j}_{k}(n)$ is the Kalman gain and $\mathbf{e}^{i,j}_{k}(n)$ is the innovation vector. The estimated channel, i.e., $\hat{h}^{i,j}_k(n)$, at time instant $n$, can be obtained from $\hat{\mathbf{x}}^{i,j}_{k}(n|n)$ given in \eqref{est_channel}. Besides, \eqref{pred_channel} gives one-step ahead prediction of the channel, where $\big [\cdot\big ]_{1}$ denotes the first element of the resultant vector. 
\par We utilize above KF as an estimator and predictor in our proposed method. The proposed method consists of an assessment phase, initialization phase, and the prediction phase. In the following subsections, we describe each in detail.
\subsection{Assessment Phase}\label{assesment_phase}
\begin{figure}
\centering
\includegraphics[scale=0.63]{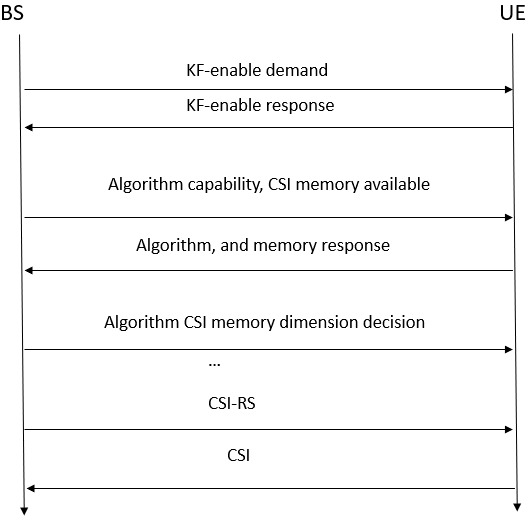}
\caption{Description of assessment phase.}
\label{algo}
\end{figure}

During the assessment phase, BS and the UE exchange messages in order to: 1) assess the available capabilities at the UE, 2) agree on the algorithm to be used (e.g., in our study, we use KF), 3) establish the amount of past CSI estimation, and 4) the length of the initialization phase. The assessment phase is summarized in Fig.\,\ref{algo}.
\subsection{Initialization Phase}
During the initialization phase (IP), the CSI acquisition procedure follows conventional method, where the UE sends only the compressed version of the estimated channel, $\hat{h}^{UE}_{k}(n)$, to the BS. Here, for brevity, we have dropped the antenna's index. Therefore, CSI received at the BS, $\hat{h}^{BS}_{k}(n)$, is given as
\begin{equation}\label{conventional}
    \hat{h}^{BS}_{k}(n)=F_{c}\{\hat{h}^{UE}_{k}(n)\}\:,
\end{equation}
where $F_{c}\{\cdot\}$ is a function that includes quantization and reduces the overhead necessary to feedback information to the BS during conventional CSI acquisition. For example, $F_{c}\{\cdot\}$ can be a standard element-wise quantization scheme or the Type-I and Type-II CSI report described in the third Generation Partnership Project (3GPP) \cite{3GPP1, 3GPP2}.
\subsection{{Prediction Phase}}
In the prediction phase, the UE and the BS utilize the KF (discussed in Section\,\ref{KF_prediction}), to predict the channel. It is, however, important to notice that since both the network entities are using the same predictor function and prediction is based on the previously known CSI estimation, thereby, the prediction at both ends would be same, i.e., $\widetilde{h}^{BS}_{k}(n-1)=\widetilde{h}^{UE}_{k}(n-1)$; where $\widetilde{h}^{BS}_{k}$ and $\widetilde{h}^{UE}_{k}$ represent the predicted channel at the BS and the UE, respectively. Nevertheless, it is expected that there will be a certain difference between the predicted and the estimated channel at the UE. Therefore, in the proposed method, the UE computes an update as
\begin{equation}\label{update}
    \Delta_{k}(n)=\varrho\{\widetilde{h}^{UE}_{k}(n-1), \hat{h}^{UE}_k(n)\}\:,
\end{equation}
where the function $\varrho\{\cdot\}$ (named the update function) is a measure of the distance between the two values. The most simple implementation of such a function is a difference, i.e., $\Delta_{k}(n)=\widetilde{h}^{UE}_{k}(n-1)- \hat{h}^{UE}_k(n)$.
Once the update is computed, it is quantized through a dedicated function $F_{Q} \{\cdot\}$ and fed back to the BS as
\begin{equation}\label{feedback}
    \Omega_{k}(n)=F_{Q}\{\Delta_{k}(n)\}\:,
\end{equation}
where a simple implementation of the function $F_{Q} \{\cdot\}$, is a quantization function. The update given in \eqref{feedback} will be reported to the BS, where the BS will estimate the channel as 
\begin{equation}\label{estimated_BS}
    \hat{h}^{BS}_k(n)=\widetilde{h}^{BS}_{k}(n-1)+\varrho^{-1}(\Omega_{k}(n))\;,
\end{equation}
where $\varrho^{-1}$ represents an appropriate inverse of the function $\varrho$. If the update function is a simple difference, the inverse will be $\varrho^{-1}=-\varrho$, therefore, the above equation can be written as 
\begin{equation}\label{Hbs}
    \hat{h}^{BS}_k(n)=\widetilde{h}^{BS}_{k}(n-1)-F_{Q}\{\widetilde{h}^{UE}_{k}(n-1)- \hat{h}^{UE}_k(n)\}\;,
\end{equation}
which gives the estimated channel at the BS. The benefit of using above approach for channel estimation is twofold. Firstly, in the above equation, if the predicted and estimated channel at the UE are the same then there is no need to feedback anything; thus, the estimated channel at the BS will only be $\hat{h}^{BS}_k(n)=\widetilde{h}^{BS}_{k}(n-1)$. In such a scenario, feedback will be eliminated, which ultimately reduces the overhead and is of supreme importance in massive MIMO environment, where the feedback requirement grows tremendously with the increase in number of antennas \cite{lim_fb}. On the other hand, if UE needs to feedback anything then the quantizer will introduce less noise due to feeding back the difference. In other words, if \eqref{update} has smaller dynamics than the conventional method (i.e., \eqref{conventional}), then the proposed method will need a fewer bits to send the feedback. Thus, BS can estimate the channel by simply calculating the difference between the predicted channel, $\widetilde{h}^{BS}_k(n-1)$, and the quantized feedback, i.e, $F_{Q}\{\widetilde{h}^{UE}_{k}(n-1)- \hat{h}^{UE}_k(n)\}$. Finally, considering the proposed method\footnote{{Here, it is important to mention that in case of reciprocity-based channel estimation, the proposed method would work if the two estimates can be guaranteed similar enough.}}, we make the following remarks.
\subsubsection{Remark 1}
The advantage of the proposed method is either to reduce the amount of bits necessary for the feedback, i.e., $\Omega_{k}(n)$ needs less bits than $F_{c}\{\hat{h}^{UE}_{k}(n)\}$ for similar performance, or to increase the performances with the same amount of bits. 
\subsubsection{Remark 2}
If the prediction was perfect, i.e., $\widetilde{h}^{UE}_{k}(n-1)=\hat{h}^{UE}_{k}(n)$, then no feedback would be necessary, bringing the amount of necessary bits to 0; thus, reaping benefit of further reducing the overhead.
\subsubsection{Remark 3}
If infinite amount of bits are considered for the feedback, it is possible to define $F_Q \{\varkappa\}=F_C \{\varkappa\}= \varkappa$, where $\varkappa$ is the quantity to be quantized. This would imply that $\hat{h}^{BS}_{k}(n)=\hat{h}^{UE}_{k}(n)$, i.e., the proposed method has no advantage. In other words, the largest amount of gain is acquired with low resolutions feedback. This is particularly relevant as 3GPP CSI acquisition schemes consider low amount of bits to reduce feedback overhead.
\subsubsection{Remark 4}
Different prediction algorithms can be tabled and standardized. Possible standardization points include: algorithm, memory, and message exchanges.
\section{Results and Discussions}\label{sec4}
 In this section, firstly, we describe the simulation environment, including system parameters, and then we present simulation results, through which the performance of our proposed method is evaluated.
\subsection{Simulation Environment}
We consider two network entities, i.e., one BS and one UE, where the former is equipped with four antennas and the latter with two. Moreover, the results are obtained using two different evaluation parameters, which are given below.
\subsection{Evaluation Parameters}
In order to assess the performance of proposed channel reporting method, two evaluation parameters, i.e., MSE and received signal-to-noise ratio (SNR) \textendash { at the UE}, are calculated by using the actual channel, ${h}_{k}(n)$, given in \eqref{channel_AR}, and the estimated channel, $\hat{h}^{BS}_{k}(n)$, at the BS. The mathematical expressions for the calculation of MSE is given below
\begin{equation}\label{MSE}
    MSE ={\frac{1}{N} }{\sum_{n=1}^{N}} \mid\mid{{h}_{k}(n)-\hat{h}^{BS}_{k}(n)}\mid\mid_{FRO}^{2}\:,
\end{equation}
where $\mid\mid \cdot \mid\mid_{FRO}$ indicates the Frobenius norm. The MSE measures the distance between the actual channel, ${h}_{k}(n)$, and the estimated channel, $\hat{h}^{BS}_{k}(n)$. Therefore, a perfect estimator would have a MSE equal to zero.
\par Furthermore, to assess the gains in terms of communication parameters, we use the estimated channel, i.e. $\hat{h}^{BS}_{k}(n)$, to compute a simple matched filter (MF) precoder, and we measure the received SNR at the UE by considering SU-MIMO scenario.
\par In the following, we show how the proposed method can either reduce the MSE at a parity of overhead bit or can provide same level of error using fewer feedback bits. Besides, the improved performance of received SNR in the proposed method is evaluated.  
\subsection{Analysis of MSE}
\begin{figure}
\centering
\includegraphics[scale=0.59]{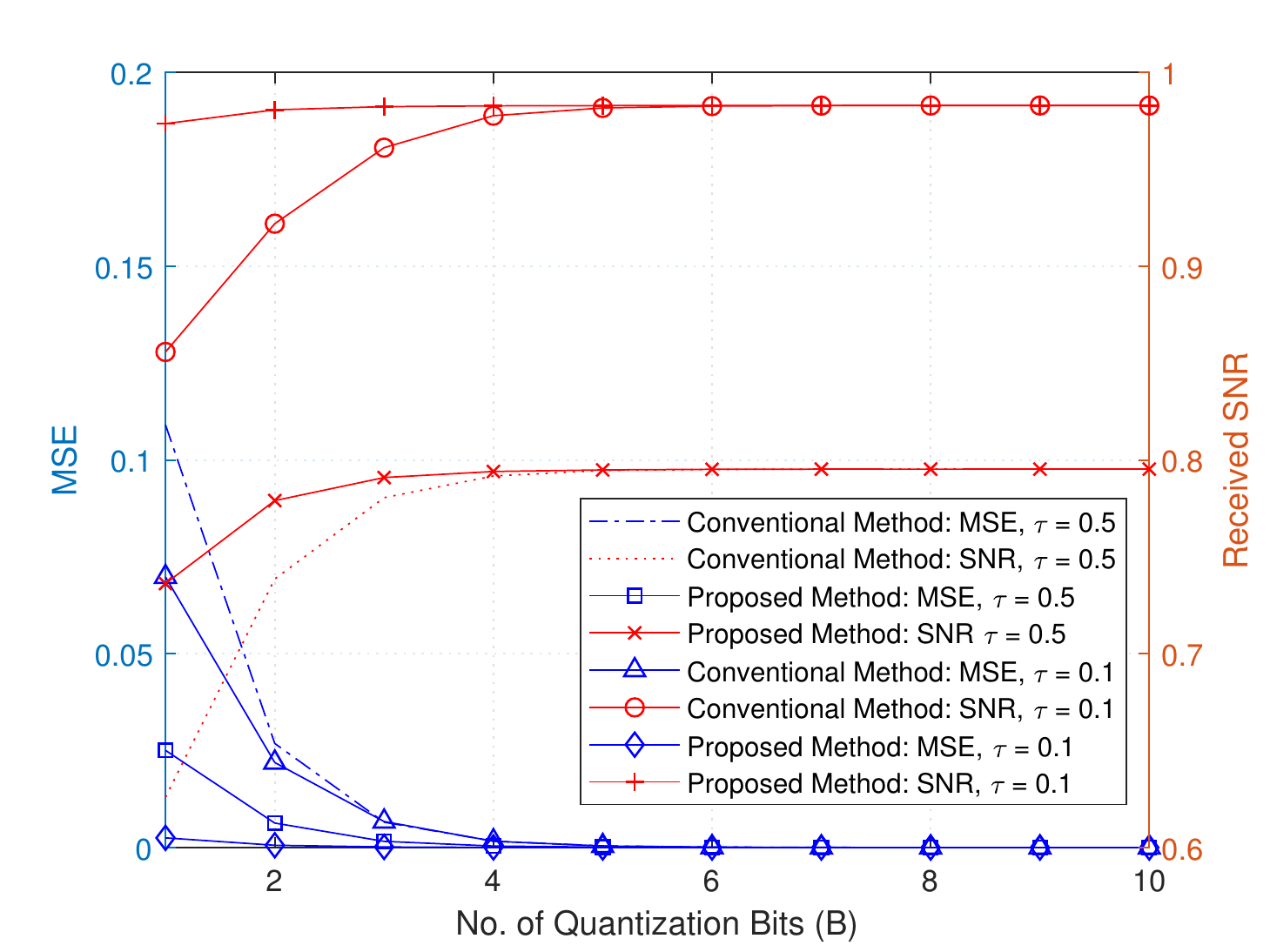}
\caption{A comparison between the number of quantization bits vs. MSE and received SNR under $\tau = {0.5,\,0.1}$.}
\label{quant}
\end{figure}

Fig.\,\ref{quant} depicts the performance of MSE under the conventional and the proposed methods, when the quantization bits, $B$, are increased from $1$ to $10$. Additionally, different curves are portrayed by considering different values of $\tau$, $\tau= 0$ indicates a channel perfectly predictable and $\tau= 1$ shows a channel for which each realization is independent of the previous. The figure reveals that increasing the quantization bits results in reducing the MSE, in both the methods, i.e., conventional and proposed. Nonetheless, the proposed method outperforms the conventional method. Additionally, it can be observed that increasing $\tau$ results in reducing less percentage of MSE, which is due to the reason of inaccurate prediction. Nevertheless, there is still a significant reduction, and the performance of the conventional method touches the proposed one when B is $4$. In a nutshell, the proposed method not only reduces the MSE but also saves the quantization bits.
\par Fig.\,\ref{tau} reveals the behaviour of MSE when the $\tau$ is increased from $0$ to $1$ and with two different quantization bits. The figure shows that the MSE increases in the proposed method with the increase of $\tau$, for both values of quantization bits. This is due to the reason of bad prediction in the low SNR regime, which results in adding more noise on both sides of the predictors; thus, increasing the MSE. However, increasing the quantization bits to $2$, results in reducing the MSE as compared to $1$-bit quantizer. Conversely, the MSE of the conventional method is reducing with a negligible margin. In summary, the performance of the proposed method is significantly better than the conventional method. Notably, the performance of the conventional method gets better than the proposed one after approximately $\tau$ is $0.92$, which is due to the reason of bad prediction in the proposed method. In other words, the channel is almost independent and identically distributed (iid); consequently, the noise will be added at both ends of the communication system, which ultimately increases the MSE in the proposed method when $\tau= 0.92$.
\subsection{Analysis of Received SNR at UE}
\par Fig.\,\ref{quant} shows a comparison between the received SNR and the quantization bits with $\tau$ equals $0.5$ and $0.1$. Its can be seen that the proposed method gives better SNR values under both the $\tau$. Also, there is no improvement in the received SNR, in both the methods, after $5$ quantization bits. This is in line with Fig.\,\ref{quant}, for which with five quantization bits the MSE was very close to zero for both methods. In short, for one quantization bit, the proposed method gives an SNR gain of $17$\% and $13$\% when $\tau$ is $0.5$ and $0.1$, respectively. This illustrates that SNR gain is good for lower $\tau$, which is understandable. It can also be observed that the proposed method saves a significant amount of quantization bits for both values of $\tau$.
\begin{figure}
\centering
\includegraphics[scale=0.59]{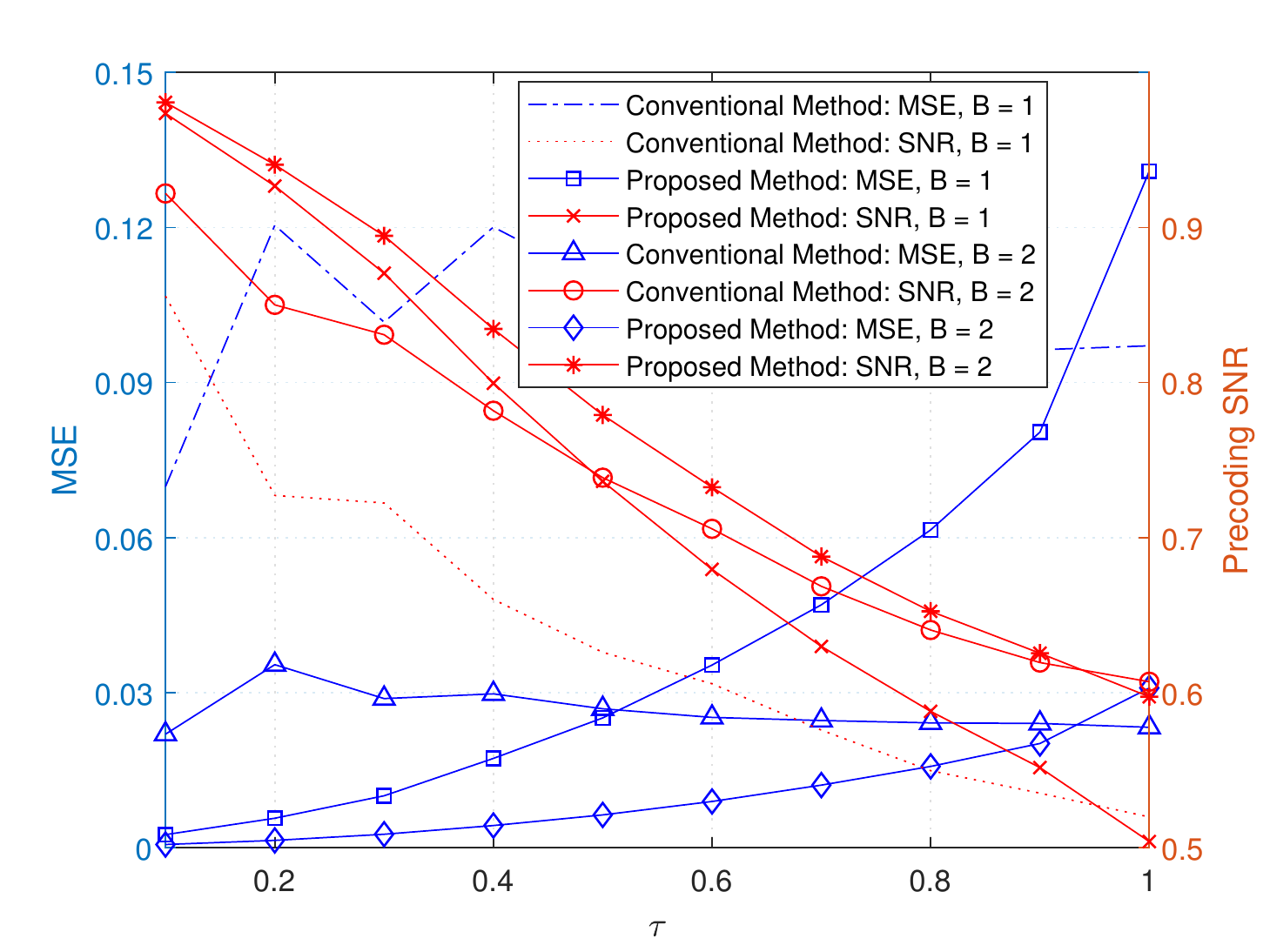}
\caption{Performance of MSE and received SNR with changing $\tau$ and under $B = {1, 2}$.}
\label{tau}
\end{figure}
\par Finally, the performance of received SNR at the UE can be observed in Fig.\,\ref{tau} when $\tau$ is changed from $0.1$ to $1$. Besides, a comparison is also drawn for different quantization bits. It can be depicted that the strength of received SNR decreases linearly when $\tau$ increases, in both methods. Nonetheless, the proposed method outperforms the conventional one. 
\par Summarily, the proposed method results in not only saving the quantization bits but also reducing the MSE and increasing the SNR gain. This shows that the proposed method can play a pivotal role in reducing the OTA overhead and providing better quality-of-service (QoS). Thus, it can be a viable solution for reporting CSI in potential 5G and beyond applications.  
\section{Conclusion} \label{conclusion}
Motivated by the proliferation of cellular users and the evolution of MIMO communication, this paper addressed a novel algorithm to report CSI in 5G and beyond wireless networks. The proposed work consists of three major steps, that is, assessment phase, initialization phase, and the prediction phase. Particularly, the work has introduced the idea of using a channel predictor function based on KF, at both ends of the communication system. The output of the predictor is used as a base, and an updated channel measure is reported with respect to the prediction. Simulation results showed a significant improvement in the proposed method, in terms of reducing overhead, MSE, and maximizing the received SNR. Therefore, the proposed work is beneficial to meet the requirements of existing and futuristic cellular networks. 
\bibliographystyle{IEEEtran}
\bibliography{main}
\end{document}